\documentclass[prl,twocolumn,aps,floatfix,10pt,showpacs,superscriptaddress]{revtex4}
\usepackage{amsmath}
\usepackage{amsfonts}
\usepackage{amssymb}
\usepackage{graphicx}
\usepackage{ucs}
\usepackage[utf8x]{inputenc}
\usepackage{psfrag,layout}
\usepackage{upgreek}
\usepackage{subfigure}

\renewcommand{\vec}[1]{\mathbf{#1}}

\begin{document}

\title{Consistent Hydrodynamics for Phase Field Crystals}

\author{V. Heinonen}
\affiliation{COMP Centre of Excellence,
Department of Applied Physics,
Aalto University, School of Science,
P.O.Box 11100,
FI-00076 Aalto
Finland}
\email{vili.heinonen@aalto.fi}
\author{C. V. Achim}
\affiliation{COMP Centre of Excellence,
Department of Applied Physics,
Aalto University, School of Science,
P.O.Box 11100,
FI-00076 Aalto
Finland}
\author{J. M. Kosterlitz}
\affiliation{Department of Physics, Brown University, 
Providence RI 02912-1843, USA}
\author{See-Chen Ying}
\affiliation{Department of Physics, Brown University, 
Providence RI 02912-1843, USA}
\author{J. Lowengrub}
\affiliation{Department of Mathematics, University of California, Irvine, CA 92697, USA}
\affiliation{Department of Chemical Engineering and Materials Science, University of California, Irvine, CA 92697, USA}
\author{T. Ala-Nissila}
\affiliation{COMP Centre of Excellence,
Department of Applied Physics,
Aalto University, School of Science,
P.O.Box 11100,
FI-00076 Aalto
Finland}
\affiliation{Department of Physics, Brown University, 
Providence RI 02912-1843, USA}

\begin{abstract}
We use the amplitude expansion in the phase field crystal framework to formulate an approach where the fields describing the microscopic structure of the material are coupled to a hydrodynamic velocity field. The model is shown to reduce to the well known macroscopic theories in appropriate limits, including compressible Navier-Stokes and wave equations. Moreover, we show that the dynamics proposed allows for long wavelength phonon modes and demonstrate the theory numerically showing that the elastic excitations in the system are relaxed through phonon emission. 
\end{abstract}


\maketitle

One of the grand challenges in materials modeling is to take into account the large range of different time scales from elastic vibrations to vacancy diffusion and length scales varying from atomistic details to dislocations and grain boundaries at micron scales. Phase field crystal (PFC) models were originally introduced \cite{Elder:2002eq} in order to couple diffusive time scales with atomistic spatial resolution and are a suitable candidate for a framework with a wide range of temporal scales. This is achieved by coarsening out fluctuations due to finite temperature by describing the system in terms of a mass density field which is averaged over thermal fluctuations. Over the past decade PFC models have been used successfully to study a wide variety of different phenomena in solids \cite{Emmerich:2012ko}. 

One of the important advantages of the PFC models is the intrinsic incorporation of elastic energy associated with a fixed inter-atomic length scale. However, this poses a great challenge for the dynamics of the system: elastic excitations emit phonons which cannot be described using over-damped, purely dissipative dynamics. An attempt to include fast time scales in the dynamics was with the introduction of an explicit second order time derivative in the equation of motion for the PFC mass density field $\tilde{\rho}$ as 
\begin{equation}
\partial_t^2 \tilde{\rho} + \alpha \partial_t \tilde{\rho} = \nabla^2 \frac{\delta \tilde{F}}{\delta \tilde{\rho}},
\label{eq:MPFC}
\end{equation}
where $\tilde{F}$ is a PFC free energy and $\alpha$ a dissipation parameter \cite{Stefanovic:2006fg,Galenko:2009hr}. The incorporation of the second order time derivative gives rise to short wavelength oscillations accelerating relaxation processes but fails to describe large scale vibrations. This was pointed out by Majaniemi \textit{et al.} who studied coupling of a displacement field to the mass density field within the PFC framework \cite{Majaniemi:2007iy,Majaniemi:2008cq}.

Fast dynamics have been studied more systematically by coupling a velocity field with the PFC mass density field \cite{Ramos:2010fa,Baskaran:2014ko}. However, two main obstacles  arise from this sort of coupling. First, the PFC mass density field oscillates at an atomistic length scale creating large gradients  which result  in spurious  unphysical flows. Second, it is not clear how dissipation at microscopic length scales should be  incorporated.  Hydrodynamics considers smooth fields and it is hard to extend the theory to spatially microscopic systems with velocity variations at the inter-atomic length scale.

Some attempts have been made recently to overcome these problems by introducing a mesoscopic mass density which can be obtained by smoothing out the PFC mass density with specific Fourier filters \cite{Toth:2013ie} and by considering colloidal systems where hydrodynamics is  solved only in the solvent surrounding the colloidal particles \cite{Praetorius:2015co}. In this Letter we introduce an approach that avoids the possible ambiguity of coarse graining the fields and that is not limited to colloidal systems.

In this work, we follow the idea of coarse-graining the mass density and velocity fields by using the amplitude expansion framework \cite{Goldenfeld:2005dq,Athreya:2006hs} where the structure is described by the amplitudes of the atomistic density oscillations instead of the PFC mass density field itself. This framework allows for a description of the material by smooth fields  and it can be shown to  reduce to well known macroscopic theories. The displacement field is naturally coupled to the amplitudes of the density oscillations and to the velocity field, with no need for additional assumptions.

We derive the dynamical equations for the system by first writing down energy conserving dynamics 
for the PFC system and then coarse-graining these equations as well as the energy in order to obtain conserved dynamics for the mesoscopic system generated by a mesoscopic energy. After this we add dissipation in the system to make the dynamics irreversible. We consider some limits of the model and study the grain rotation problem to make a connection with relaxing elastic excitations through phonon emission. 

\paragraph{Conserved dynamics and coarse-graining.}

We start by writing down conserved dynamics generated by an effective Hamiltonian
\begin{equation}
\tilde{\mathcal{H}}[\tilde{\rho},\tilde{\vec{v}}] = T[\tilde{\rho},\tilde{\vec{v}}] + \tilde{F}[\tilde{\rho}],
\end{equation}
where $T[\tilde{\rho}, \tilde{\vec{v}}]=\int d\vec{r} (\tilde{\rho} |\tilde{\vec{v}}|^2/2)$ is the kinetic energy and $\tilde{F}[\tilde{\rho}]$ is any configuration free energy of the PFC type with a periodic ground state in the solid phase. Here $\tilde{\rho}$, $\tilde{\vec{v}}$ are the PFC mass density and velocity fields, respectively. We assume conservation of mass and momentum density given by
\begin{align}
\partial_t \tilde{\rho} &= -\nabla \cdot (\tilde{\rho} \tilde{\vec{v}}),
\label{eq:micro_density_conservation} \\
\partial_t (\tilde{\rho} \tilde{\vec{v}}) &= - \nabla \cdot (\tilde{\rho} \tilde{\vec{v}} \otimes \tilde{\vec{v}}) + \tilde{\vec{f}},
\label{eq:micro_momentum_conservation}
\end{align}
where $\tilde{\vec{f}}$ is a force term determined by total energy conservation.

We expand the density $\tilde{\rho}$ in Fourier space as
\begin{equation}
\tilde{\rho}(\vec{r},t) \approx \rho(\vec{r},t) + \sum_j \left[
\eta_j(\vec{r},t) e^{i \vec{q}_j \cdot \vec{r}} + \text{C.C.}
\right]
\label{eq:microscopic_density}
\end{equation}
Here $\vec{q}_j$ are the reciprocal lattice vectors, $\eta_j$ are the amplitudes, $\rho$ is the  density field averaged over a unit cell of the Bravais lattice and $\text{C.C.}$ denotes the complex conjugate. 

The amplitudes $\eta_j$ and the density $\rho$ are assumed to be slowly varying in space and are treated as constants over a length scale $1/|\vec{q}_j|$. Furthermore, the amplitudes $\eta_j$ are taken to be complex valued to allow for  displacements. Change of coordinates $\vec{r} \to \vec{r} - \vec{u}(\vec{r})$ in Eq.~\eqref{eq:microscopic_density}, where $\vec{u}$ is a \emph{spatially slowly varying} displacement field results in $\eta_j \to \eta_j \exp{(-i \vec{q}_j \cdot \vec{u})}$ giving a meaning to the phase of the complex amplitudes.

Following Ref.~\cite{Athreya:2006hs} we coarse-grain Eqs.~\eqref{eq:micro_density_conservation} and 
\eqref{eq:micro_momentum_conservation}
to obtain time-evolution equations for fields $\eta_j$, $\rho$ and a mesoscopic velocity $\vec{v}$. We present the results here, the details may be found in the Supplementary Material \cite{supplementary}.

From the mass density conservation \eqref{eq:micro_density_conservation} we get
\begin{align}
\partial_t \rho &= - \nabla \cdot (\rho \vec{v}),
\label{eq:rhoeq_conserved} \\
\partial_t \eta_j &= -\mathcal{Q}_j \cdot (\eta_j \vec{v}),
\label{eq:etaeq_conserved}
\end{align}
where $\mathcal{Q}_j=\nabla + i \vec{q}_j$. The momentum density conservation of Eq.~\eqref{eq:micro_momentum_conservation} gives
\begin{equation}
\rho \frac{D \vec{v}}{Dt} := \rho \left(\partial_t \vec{v} + \vec{v} \cdot \nabla \vec{v}\right) = \vec{f},
\label{eq:veloeq_conserved}
\end{equation}
for the mesoscopic velocity with the help of Eq.~\eqref{eq:rhoeq_conserved}. 

The mesoscopic force term $\vec{f}$ in Eq.~\eqref{eq:veloeq_conserved} is determined by the conservation of the effective Hamiltonian $\mathcal{H} = T[\rho,\vec{v}] + F[\rho,\lbrace \eta_j \rbrace]$, where $T$ is the kinetic energy 
\begin{equation}
T = \int d \vec{r} \left(
\frac{1}{2} \rho |\vec{v}|^2
\right),
\label{eq:kinetic_energy}
\end{equation}
and $F$ is a configuration free energy obtained from coarse-graining $\tilde{F}$ and described in terms of $\rho$ and $\lbrace \eta_j \rbrace$. We require that $\partial_t \mathcal{H} = 0$. This results in
\begin{equation}
\begin{split}
 \vec{f} &=
- \rho \nabla \frac{\delta F}{\delta \rho} 
-  \sum_j
\left[  \eta_j^* \mathcal{Q}_j 
\frac{\delta F}{\delta \eta_j^*}  
 + \textrm{C.C.}
\right].
\end{split}
\label{eq:force}
\end{equation}

For the remainder of this article we choose a configuration free energy of a 2D hexagonal lattice
\begin{equation}
\begin{split}
&F = \int d\vec{r} \left[ \vphantom{\sum_{j=1}^3}
\frac{B^{\ell}}{2} \rho^2 -\frac{\tau}{3} \rho^3
+ \frac{\nu}{4} \rho^4 
+ \frac{\tilde{B}^x}{2}|\nabla \rho|^2  \right. \\ &\left.
+ \left( \frac{\Delta B}{2} - \tau \rho  +\frac{3\nu}{2} \rho^2 \right) A^2
+ \sum_{j=1}^3 B^x |\mathcal{G}_j \eta_j|^2  \right. \\ &\left.
+ \left( 6 \nu \rho -2\tau \right)  \left( \prod_{j=1}^3 \eta_j + \textrm{C.C.} \right) 
\right. \\ &\left.
+\frac{3\nu}{4} \left( 
A^4  
- 2 \sum_{j=1}^3 |\eta_j|^4 \right)
\right],
\end{split}
\label{eq:internal_energy}
\end{equation}
where $A^2 = 2\sum_j |\eta_j|^2$, $\mathcal{G}_j = \nabla^2 + 2i \vec{q}_j \cdot \nabla$, $B^{\ell} =\Delta B + B^x$, $\tau$ and $\nu$ are bulk energy parameters
and $\tilde{B}^x$ is a surface energy parameter for the density $\rho$. We have chosen a representation for the vectors $\vec{q}_j$ as $\vec{q}_1=(-\sqrt{3}/2,-1/2)$, $\vec{q}_2=(0,1)$ and $\vec{q}_3=(\sqrt{3}/2,-1/2)$. This energy can be obtained from the standard PFC free energy
\begin{equation*}
\tilde{F} = \int d{\bf r}\left[
\frac{\Delta B}{2}\tilde\rho^{2} 
+ \frac{B^{x}}{2}\tilde\rho(1+\nabla^{2})^{2}\tilde\rho 
- \frac{\tau}{3}\tilde\rho^{3}
+\frac{\nu}{4}\tilde\rho^{4} \right]
\label{eq:PFC_free_energy}
\end{equation*}
by coarse-graining, as discussed in \cite{Yeon:2010fz}.

Using the configuration free energy $F$ the functional derivatives in Eq.~\eqref{eq:force} become
\begin{equation}
\begin{split}
\frac{\delta F}{\delta \eta_j^* } &= 
(\Delta B - 2\tau \rho + 3\nu \rho^2) \eta_j 
+ B^x \mathcal{G}_j^2 \eta_j \\ &
+ (6 \nu \rho -2 \tau) \prod_{i \neq j} \eta_i^*  
+ 3\nu(A^2 - |\eta_j|^2) \eta_j 
\end{split}
\label{eq:deta}
\end{equation}
and
\begin{equation}
\begin{split}
\frac{\delta F}{\delta \rho } &= 
\left(  B^{\ell} + 3 \nu A^2 - \tilde{B}^x \nabla^2 \right) \rho
-\tau \rho^2 \\ &
- \tau A^2 
+\nu \rho^3  
+6\nu \left( \eta_1 \eta_2 \eta_3 + \text{C.C.} \right).
\end{split}
\label{eq:drho}
\end{equation}
%
%We have derived conserved dynamics generated by the effective Hamiltonian $\mathcal{H}$ as described by Eqs.~\eqref{eq:rhoeq_conserved}, \eqref{eq:etaeq_conserved} and \eqref{eq:veloeq_conserved}. For details on the derivation see Supplementary Material \cite{supplementary}. 

\paragraph{Dissipation.}

To incorporate irreversible effects in the dynamics we add dissipation. For the time evolution of the velocity we choose Navier-Stokes type dissipation resulting in
\begin{equation}
\rho \frac{D \vec{v}}{Dt} = \vec{f} + \mu_S \nabla^2 \vec{v} + (\mu_B - \mu_S) \nabla \nabla \cdot \vec{v},
\label{eq:velocity_nonconserved}
\end{equation}
where $\mu_S$ is a surface dissipation parameter and $\mu_B$ accounts for bulk dissipation. Source terms in the time evolution of the complex amplitude and density provide additional modes of dissipation:
\begin{align}
\partial_t \eta_j &= -\mathcal{Q}_j \cdot (\eta_j \vec{v})  - \mu_{\eta} \frac{\delta \mathcal{H}}{\delta \eta^*},
\label{eq:eta_nonconserved} \\
\partial_t \rho &= - \nabla \cdot (\rho \vec{v}) + \mu_{\rho} \nabla^2 \frac{\delta \mathcal{H}}{\delta \rho}.
\label{eq:rho_nonconserved}
\end{align}
Here $\mu_{\eta}$ and $\mu_{\rho}$ are dissipation parameters. 

Now we have complete dynamics for the system determined by Eqs.~\eqref{eq:velocity_nonconserved}, \eqref{eq:eta_nonconserved} and \eqref{eq:rho_nonconserved} and it can be shown that the dynamics leads into a non-increasing effective Hamiltonian $\mathcal{H}$ in time \cite{supplementary}. 
Next we will study some important limits of the theory. 

\paragraph{Liquid limit.}

A liquid is described by setting the complex amplitudes $\eta_j \to 0$. In this limit, the time evolution equation for the velocity field becomes
\begin{equation}
\begin{split}
\rho \frac{D \vec{v}}{Dt} &= -\nabla \left( \frac{B^{\ell}}{2} \rho^2 -\frac{2\tau}{3} \rho^3 +\frac{3\nu}{4} \rho^4 \right)  \\ &
+ \mu_S \nabla^2 \vec{v} + (\mu_B - \mu_S) \nabla (\nabla \cdot \vec{v}),
\end{split}
\label{eq:N-S}
\end{equation}
which is accompanied by Eq.~\eqref{eq:rho_nonconserved}. When the density dissipation parameter $\mu_{\rho}\rightarrow 0$,  this pair of equations become the Navier-Stokes equations for a compressible flow where the pressure $P=\frac{B^{\ell}}{2} \rho^2 -\frac{2\tau}{3} \rho^3 +\frac{3\nu}{4} \rho^4 $ is described in terms of a virial expansion in $\rho$.  Here we take the long wavelength limit and discard any derivatives of $\rho$ of higher order than two. This also removes the dissipation in Eq.~\eqref{eq:rho_nonconserved}.

\paragraph{Over-damped limit.}

In the limit where $\mu_{\eta}$ and $\mu_{\rho}$ are large, the set of equations reduces into the usual over-damped amplitude expansion model \cite{Yeon:2010fz} described by
\begin{align}
\partial_t \eta_j &= - \mu_{\eta} \frac{\delta F}{\delta \eta_j^*},
\label{eq:overdampedeta} \\
\partial_t \rho &= \mu_{\rho} \nabla^2 \frac{\delta F}{\delta \rho}.
\label{eq:overdampedrho}
\end{align}
This limit is achieved also when the dissipation of the velocity is large \cite{supplementary}.

\paragraph{Small displacement limit.}

\begin{figure}
\includegraphics[width = \columnwidth]{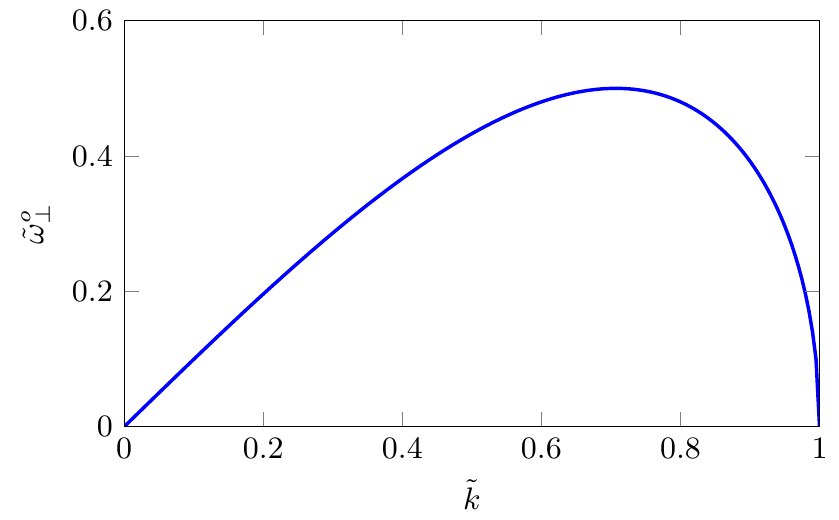}
\caption{The dispersion relation $\tilde{\omega}_{\perp}^o(\tilde{k}) = \tilde{k} \sqrt{1-\tilde{k}^2}$ for the oscillating component of the transversal wave in the small displacement limit. 
Here $\tilde{\omega}_{\perp}^o = \omega_{\perp}^o \rho_0 \mu_{\eta}/(6 \phi_0^2) $ 
and $\tilde{k}^2 = k^2 B^x \mu_{\eta}^2 \rho_0/(12 \phi_0^2) $. See text for details.}
\label{pic:dispersion_relation}
\end{figure}

Another interesting limit is the limit of small displacements. Writing the complex amplitudes as $\eta_j = \phi_j \exp{(-i \vec{q}_j \cdot \vec{u} )}$ we can rewrite the system in terms of the order parameter fields $\phi_j$ and the displacement field $\vec{u}$. Assuming a perfect hexagonal crystal, implies that  $\phi_j=\phi$. Now Eq.~\eqref{eq:eta_nonconserved} gives
\begin{align}
\partial_t \phi &= - \nabla \cdot (\phi \vec{v}) - \frac{1}{2}\mu_{\eta}  \frac{\delta F}{\delta \phi}, \\
\frac{D \vec{u}}{Dt} &= \vec{v} - \frac{1}{2} \mu_{\eta} \phi^{-2} \frac{\delta F}{\delta \vec{u}},
\end{align}
with the advective derivative $\frac{D}{Dt}$. With small enough displacements $\vec{u}$ we assume that $\rho$ and $\phi$ are constant ($\rho_0$ and $\phi_0$) and keep only $\vec{u}$ and $\vec{v}$ up to linear order. Furthermore, we assume that $\vec{u}$ changes relatively slowly in space discarding all the derivatives of order higher than two. We obtain
\begin{align}
\rho_0 \partial_t \vec{v} &= \vec{f} \approx -\frac{\delta F}{\delta \vec{u}},
\label{eq:velosmalldisplacement}\\
\partial_t \vec{u} &= \vec{v} - \frac{1}{2} \mu_{\eta} \phi^{-2} \frac{\delta F}{\delta \vec{u}},
\label{eq:usmalldisplacement} \\
\frac{\delta F}{\delta \vec{u}} &\approx - 3 B^x \phi_0^2 (\nabla^2 \vec{u} + 2 \nabla \nabla \cdot \vec{u}).
\end{align}
Here we assume that there is no dissipation of velocity. Differentiating Eq.~\eqref{eq:usmalldisplacement} yields
\begin{equation}
\begin{split}
\partial_t^2 \vec{u} &= 3 B^x \phi_0^2 \rho_0^{-1} (\nabla^2 \vec{u} + 2 \nabla \nabla \cdot \vec{u}) \\ &
+  B^x \mu_{\eta} \partial_t (\nabla^2 \vec{u} + 2 \nabla \nabla \cdot \vec{u}),
\end{split}
\label{eq:waveeq}
\end{equation}
by substituting $\partial_t \vec{v}$ from Eq.~\eqref{eq:velosmalldisplacement} giving us a damped wave equation for the hexagonal crystal symmetry. 

With the ansatz $\vec{u}={\rm exp}\left(i\vec{k}\cdot\vec{r}-\omega t\right)$, we find the dispersion relation $\omega=\omega(\vec{k})$. In particular, for the transverse modes $\vec{u}=\vec{u}_{\perp}$ with $\vec{k}\cdot\vec{u}_{\perp}=0$, we obtain
\begin{equation}
\omega_{\perp}^2 -  B^x k^2 \mu_{\eta} \omega_{\perp} + 3 B^x \phi_0^2 k^2 \rho_0^{-1} = 0,
\end{equation}
which we can solve for $\omega_{\perp} = \omega_{\perp}^d + i \omega_{\perp}^o$ giving
\begin{align}
\omega_{\perp}^d &=\frac{1}{2} B^x  \mu_{\eta} k^2, \\
\omega_{\perp}^o  &= \pm \frac{k}{2} \sqrt{\frac{B^x}{\rho_0} (12 \phi_0^2 -  B^x  \mu_{\eta}^2 \rho_0 k^2 )},
\end{align}
if $k^2 < 12 \phi_0^2/( B^x \mu_{\eta}^2 \rho_0 )$. Here $\omega_{\perp}^d$ is the damping component and $\omega_{\perp}^o$ is the oscillating component. If $k^2 > 12 \phi_0^2/( B^x \mu_{\eta}^2 \rho_0 )$ we get pure damping with
\begin{equation}
\omega_{\perp}^d =\frac{1}{2} B^x  \mu_{\eta} k^2 \pm \frac{k}{2} \sqrt{\frac{B^x}{\rho_0} (  B^x  \mu_{\eta}^2 \rho_0 k^2 - 12  \phi_0^2)},
\end{equation}
where the complete solution is a superposition of these two modes. 

Fig.~\ref{pic:dispersion_relation} shows the dispersion relation for the oscillating component in the damping and oscillating cases. 
Our result shows that in the long wave length limit, the oscillating small displacement
modes correspond to propagating phonons. For wave lengths below a critical value, the modes become  purely diffusive. This is in contrast to previous studies \cite{Stefanovic:2006fg,Galenko:2009hr} using Eq.~\eqref{eq:MPFC}, where only diffusive modes exist in the long wave length limit \cite{Majaniemi:2008cq}. Note that,  when $\mu_{\eta}= 0$, the damping vanishes resulting in an energy conserving wave equation with longitudinal and transverse modes with velocities $c_{t}^{2}=3B^x \phi_0^2\rho_0^{-1}$ for transverse and $c_{l}^{2}=9B^x \phi_0^2\rho_0^{-1}$ for longitudinal modes.

\paragraph{Grain rotation.}

To test the theory numerically, we study the dynamics of a rotated circular crystalline grain embedded in a crystalline matrix. Although experimental studies of polycrystalline patterns suggest that smaller grains usually disappear at the boundary of two larger grains rather than in the middle of a single matrix \cite{Harrison2004}, the rotated grain remains important for understanding grain boundary motion and has been studied theoretically \cite{Cahn2004} using Molecular Dynamics simulations \cite{Upmanyu2006,Trautt2012} and PFC models \cite{Wu2012,Adland:2013ew}.

The rotation of the grain forms a grain boundary at the perimeter of the grain. Taken that the grain boundary motion is curvature driven it is expected that the area of the grain decreases linearly in time as the rotation angle increases \cite{Adland:2013ew}. The increase of the rotation angle is due to the conservation of dislocation cores whose number is proportional to $\gamma(t) R(t)$, where $\gamma$ is the misorientation angle and $R$ is the radius of the grain. In our calculations we fixed the energy parameters and varied the velocity dissipation parameter $\mu_S$ keeping it equal to $\mu_B$.

\begin{figure}
%\center
\subfigure[]
{\includegraphics[width=4cm]{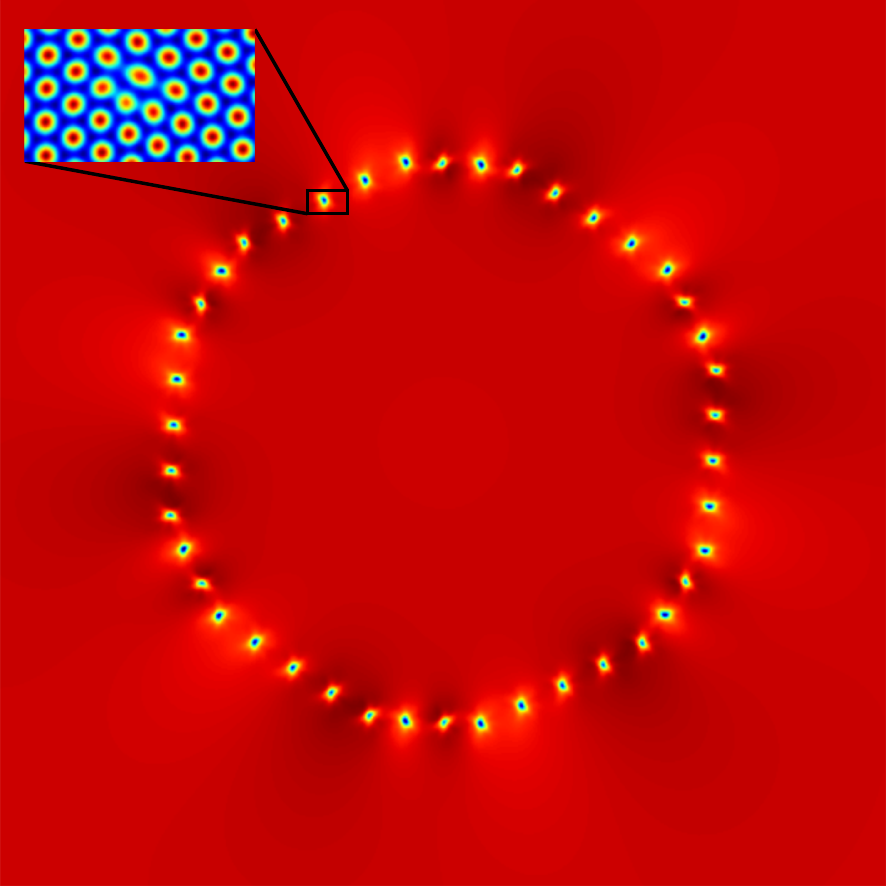}}\;
\subfigure[]
{\includegraphics[width=4cm]{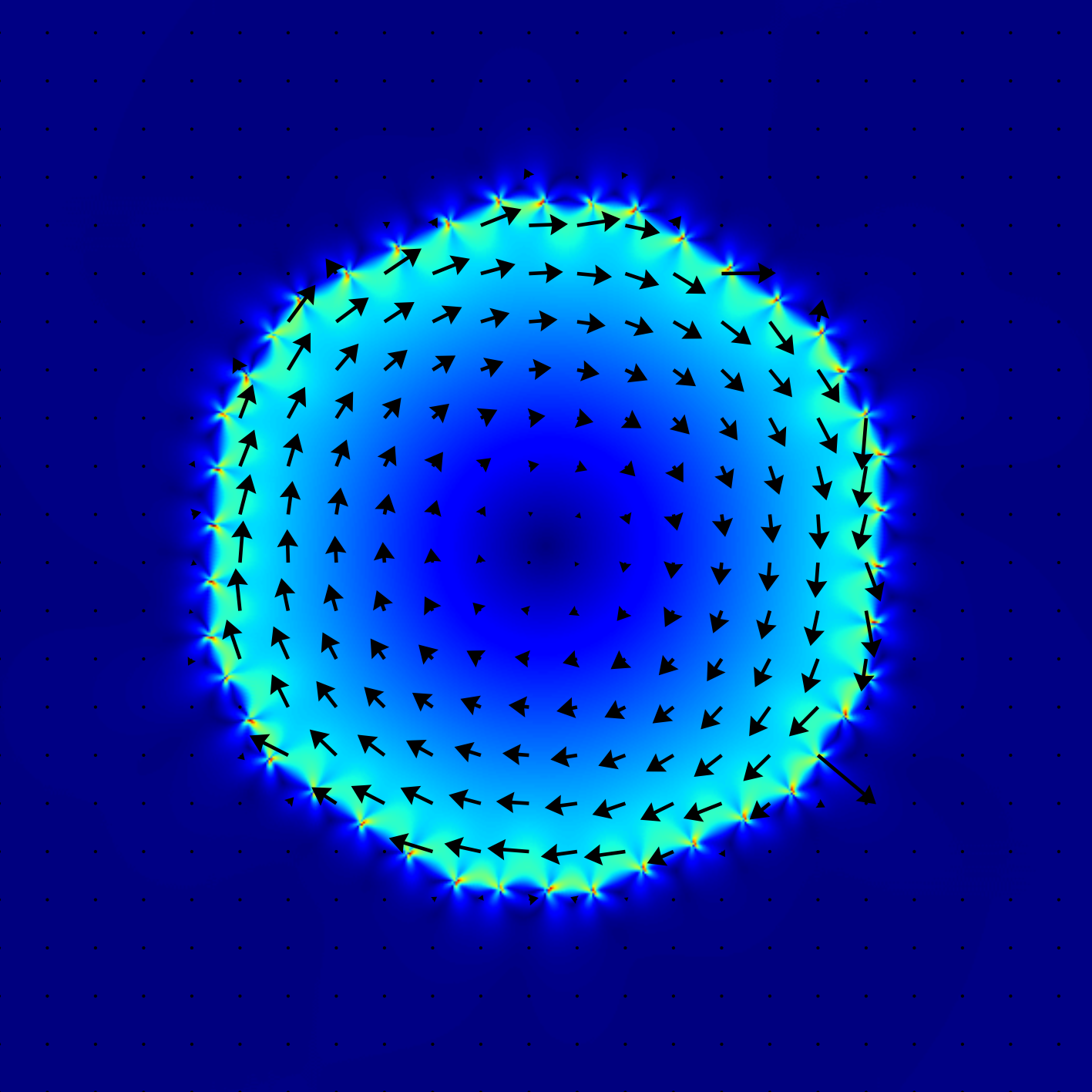}}
\caption{Panel (a) shows the density field $\rho$ with a blow-up of the reconstructed PFC density field $\tilde{\rho}$ while panel (b) shows the magnitude of the velocity field $|\vec{v}|$ with a quiver field on top to show the direction of $\vec{v}$. 
}
\label{pic:grain}
\end{figure}
Figure \ref{pic:grain} shows the density field $\rho$ and the velocity field $\vec{v}$ during the shrinking process exposing the dislocation cores at the boundary of the grain and showing the rotation of the grain facilitated by the velocity field $\vec{v}$. Note that the slowly varying  density $\rho$ does not vary much even at dislocation cores. 

\begin{figure}
\includegraphics[width=0.45\textwidth] {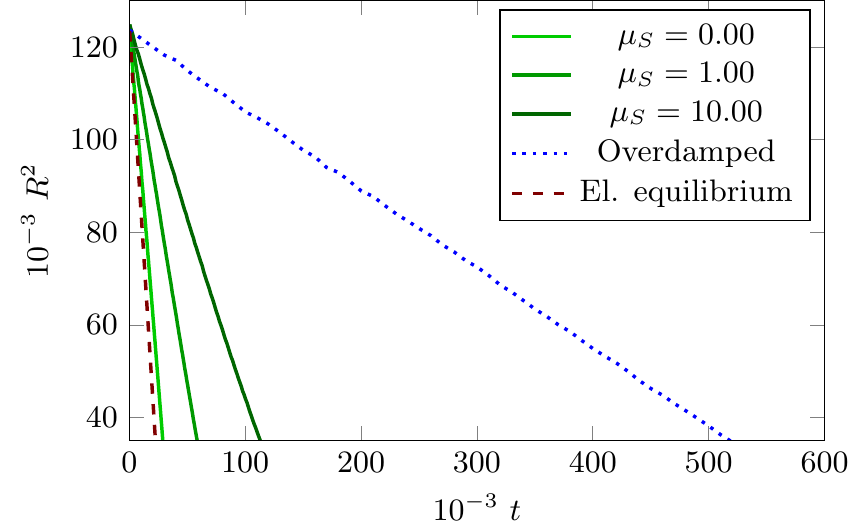}
\caption{The squared radius $R^2$ of the grain as function of time for various different values of $\mu_S$ and for over-damped dynamics and elastically equilibrated over-damped dynamics.}
\label{pic:radii}
\end{figure}

The rate of shrinking is  shown in Fig.~\ref{pic:radii}. The shrinking of the grain and also the energy dissipation is faster when we decrease the dissipation parameter $\mu_S$. For comparison we have included a calculation with over-damped dynamics given by Eqs.~\eqref{eq:overdampedeta} and \eqref{eq:overdampedrho} and also over-damped dynamics with elastic equilibration, where the energy is minimized with respect to the deformation field $\vec{u}$ at all times as described in \cite{Heinonen:2014gr}. 

Changing $\mu_S$ changes the rate of the dynamics. The dynamics in the $\mu_{S} \to 0$ limit is very similar to dynamics subject to the constraint of elastic equilibrium and we suspect the fast dynamics when $\mu_S \to 0$  is caused by the minimization of elastic excitations by creation of vibrations  which are present throughout the shrinking process  with hydrodynamics. 
We keep $\rho$ constant for  overdamped dynamics with and without elastic equilibration since the effect of  density  is negligible  in the absence of hydrodynamics. For numerical details see \footnote{The parameters used for the grain rotation calculation are $B^x=\tilde{B}^x=1$, $\Delta B = 0.097$, $\mu_{\rho} = 0.05$, $\mu_{\eta} = 1$, $\tau=0.885$, $\nu = 1$. For the spatial discretization we used $\Delta x = \Delta y = 2$ with a numerical grid of $768 \times 768$ while for the temporal discretization we used a forward Euler method with a time step of $\Delta t = 0.125$. For more details see Supplementary Material.}.

\paragraph{Summary and discussion.}

We introduce a scheme  which couples fast dynamics to dissipative processes on a mesoscopic length scale. The dynamics arises from conservation laws  which couple a velocity field with the fields describing the structure of the system in a consistent manner. We have also shown with a numerical example how the dynamics changes due to the presence of vibrating modes.

The method presented here allows for different types of dissipation in the time evolution of the system. For example, instead of the Navier-Stokes type dissipation used here,  one could  use a Langevin type dissipation $-\mu_L \vec{v}$ in the velocity equation \eqref{eq:velocity_nonconserved}. This breaks the Galilean invariance  of the velocity equation and  introduces dissipation similar to commonly used in PFC dynamics as shown by linearizing hydrodynamics \cite{Ramos:2010fa}. The Navier-Stokes type dissipation used here  avoids the problem of bulk dissipation described in \cite{Adland:2013ew} since the velocity is Galilean invariant allowing for parallel transport of all fields while the dissipation takes place only when $\nabla^2 \vec{v}\neq 0$ so that uniform motion does not dissipate energy. This also suggests that large grains are more sluggish with traditional PFC dynamics described by Eq.~\eqref{eq:MPFC} than with the full hydrodynamics since dissipation happens everywhere in the grain rather than just at the perimeter.

The approach of this letter  is general and  can be extended to any configuration free energy $F$  which can be  written in terms of slowly varying complex amplitudes and density field. We expect this approach to be useful for  problems where lattice vibrations, mass transport and other fast phenomena are coupled to the solid-liquid symmetry breaking. Some examples  of such problems are fracture dynamics, fast solidification and coarsening of polycrystalline patterns.

This work has been supported in part by the Academy of Finland through its COMP CoE grant No. 251748 and by FP7 IRSES 247504. JL would like to acknowledge partial support from the National Science
Foundation (NSF) Division
of Mathematical Sciences and from the NSF Division of Materials Research. We acknowledge the computational resources provided by the Aalto Science-IT project. The authors wish to acknowledge CSC – IT Center for Science, Finland, for generous computational resources.
We thank Zhi-Feng Huang for helpful discussions. 

\bibliography{hydrodynamics}

\end{document}